\documentclass[twocolumn,showpacs,preprintnumbers,amsmath,amssymb]{revtex4}
\usepackage{epsfig}
\usepackage{amsmath}
\usepackage{graphicx}
\usepackage{amsfonts}
\usepackage{amssymb}
\newcommand{\beq}{\begin{equation}}
\newcommand{\eeq}{\end{equation}}

\begin{document}

\title{Quasilinear approach to summation of the WKB series}

\author{V.~B.\ Mandelzweig}

\affiliation {~Racah Institute of Physics, The Hebrew University,
    Jerusalem 91904, Israel}

\begin{abstract}
\smallskip 
It is shown that the quasilinearization method (QLM) sums the WKB 
series. The method approaches solution of the Riccati equation 
(obtained by casting the Schr\"{o}dinger equation in a nonlinear 
form) by approximating the nonlinear terms 
by a sequence of the linear ones, and is not based
on the existence of a smallness parameter. Each p-th QLM iterate is 
expressible in a closed integral form. Its expansion 
in powers of $\hbar$ reproduces the structure of the WKB 
series generating an infinite number of the WKB terms. 
Coefficients of the first $2^p$ terms of the expansion are 
exact while coefficients of a similar number 
of the next terms are approximate. 
The quantization condition in any QLM iteration, 
including the first, leads to exact energies for many well 
known  physical potentials such as the Coulomb, harmonic 
oscillator, P\"{o}schl-Teller, Hulthen, Hyleraas, Morse, Eckart etc. 

\end{abstract}
\maketitle

\section{Introduction} 

The derivation of the WKB solution starts by casting the radial
Schr\"{o}dinger equation into nonlinear Riccati form and solving that
equation by expansion in powers of $\hbar$. It is interesting instead to
solve this nonlinear equation with the help of the quasilinearization
method (QLM) \cite{VBM1,VBM2,MT,KM1,KM2,K} and compare with the WKB results.
The quasilinearization method and its iterations were constructed 
as a generalization of the Newton-Raphson
 method \cite{CB,RR} to 
yield
 rapid quadratic and often monotonic convergence to the 
exact solution. It does not rely
on the existence of any kind of smallness parameter. 
The initial comparison of QLM and WKB was performed 
in the work \cite{RV,J}, where it was shown that the first QLM iteration
reproduces the structure of the WKB series generating an infinite series of
the WKB terms, but with different coefficients. Besides being a better
approximation than the usual WKB, the first QLM iteration is 
also expressible in a closed integral form.

The goal of this work is to point out that the similar conclusions 
could be reached for higher QLM approximations as well. Namely, we show that  
the $p$-th QLM iterate with $p>1$  reproduce the WKB series  exactly up 
to $\hbar^{2^p}$: when expanded in powers of $\hbar$, it, besides 
providing the correct structure of the whole series, generates the coefficients
of the first $2^n$ terms of the WKB series precisely and of a similar number
of the next terms approximately. In addition we prove that the exact 
quantization condition in any QLM iteration, 
including the first, leads to exact energies not only for the the Coulomb 
and harmonic oscillator potentials
as it was shown in ref. \cite{RV,J}, but for many other well 
known  physical potentials used in molecular and nuclear physics such as 
the P\"{o}schl-Teller, Hulthen, Hyleraas, Morse, Eckart etc.

The paper is arranged as follows: in the second chapter we cast the radial
Schr\"{o}dinger equation into the nonlinear Riccati form and solve it 
by QLM iterations. Then in the third chapter we expand the first, 
second and third QLM 
iterates  in the series in powers of $\hbar$ and and compare these 
expansions  with the usual WKB series. In the next, forth chapter,  
we apply our method to equations with
 the  Coulomb, harmonic oscillator, P\"{o}schl-Teller, Hulthen, Hyleraas, 
 Morse and Eckart potentials and prove that the QLM method provides exact 
 energies for both ground and excited  states already in the first iteration. 
 Our results, advantages of the quasilinearization method over WKB 
 and its possible future applications
are discussed in the final, fifth chapter.
\section{QLM approach to the solution of the Schr\"{o}dinger equation}

The usual WKB substitution 
\beq 
\chi(r)=C \exp\left(\lambda\int^r y(r') dr'\right),
\label{eq:ueq} 
\eeq 
converts the radial Schr\"{o}dinger equation
\beq 
\frac{d^2\chi(r)}{dr^2}+\lambda^2 k^2 \chi(r)=0 
\label{eq:seq} 
\eeq 
to nonlinear Riccati form 
\beq 
\frac{dy(z)}{dz}+ (k^2(z)+y^2(z))=0. 
\label{eq:weq} 
\eeq 
Here $k^2(z)=E-V-{l(l+1)}/{z^2}$, $\lambda^2={2 m}/{\hbar^2}$ and $z=\lambda r$.

The proper bound state boundary condition for potentials falling off at $z
\simeq z_0 \simeq \infty$ is $y(z)= \mathrm{const}$ at $z \geq z_0$.
 This means that 
\beq
y'(z_0) = 0, 
\label{eq:ypsikeq}
\eeq
so that Eq. (\ref{eq:weq}) at $z \simeq z_0$ reduces to
$k(z_0)^2+y^2(z_0))=0$ or $y(z_0))= \pm i k(z_0)$. We choose here  
to define the boundary condition with the plus sign, so that 
\beq
y(z_0)=  i k(z_0).
\label{eq:bceq} 
\eeq

The
quasilinearization \cite{VBM1,MT,RV} of this equation gives a set of 
recurrence differential equations 
\beq
\frac{dy_{p}(z)}{dz}=y_{p-1}^2(z)-2 y_{p}(z)y_{p-1}(z)-k^2(z). 
\label{eq:qeq} 
\eeq
with the boundary condition deduced from Eq. (\ref{eq:bceq}):
\beq
y_p(z_0)=  i k(z_0).
\label{eq:ppeq} 
\eeq

The analytic solution \cite{RV} of these equations expresses the $p$-th
iterate $ y_{p}(z)$ in terms of the previous iterate:
\begin{eqnarray}
 y_{p}(z)&=& f_{p-1}(z)-\int_{z_0}^{z}ds \frac{d\,f_{p-1}(s)}{ds}\;
\exp[-2 \int_{s}^{z} y_{p-1}(t) dt],  \nonumber \\
\label{eq:peq}
\end{eqnarray}
where
\begin{eqnarray}
 f_{p-1}(z)&=&
\frac{y_{p-1}^2(z)-k^2(z)}{2 y_{p-1}(z)}.
\label{eq:ipqeq}
\end{eqnarray}
Indeed, differentiation of both parts of Eq. (\ref{eq:ipqeq}) leads
 immediately to 
Eq.\ (\ref{eq:qeq}) which proves that  $y_{p}(z)$ is a solution of 
this equation. The boundary condition (\ref{eq:bceq}) is obviously 
satisfied automatically. \\
The second term in Eq. (\ref{eq:peq}) could be written as
\begin{eqnarray}
 \int_{z_0}^{z}ds \left(-\frac{\frac{d\,f_{p-1}(s)}{ds}}{2 y_{p-1}(s)}\right)\;
 \left(2 y_{p-1}(s)\;\exp[-2 \int_{s}^{z} y_{p-1}(t)\right) dt],  \nonumber \\
\label{eq:p2eq}
\end{eqnarray}
The second expression in the round brackets in the integrand is the derivative
of the exponential there. The integration by parts of this integral therefore gives:
\begin{widetext}
\begin{eqnarray} 
\left[\left(-\frac{1}{2 y_{p-1}(s)}\frac{d\,f_{p-1}(s)}{ds}\right)\,
\left.\exp[-2 \int_{s}^{z} y_{p-1}(t)]\right]\right|_{z_0}^z 
 -\int_{z_0}^{z}ds \frac{d}{ds}\left(-\frac{\frac{d\,f_{p-1}(s)}{ds}}{2 y_{p-1}(s)}
 \right)\;\exp[-2 \int_{s}^{z} y_{p-1}(t)) dt],  \nonumber \\
\label{eq:pppeq}
\end{eqnarray}
\end{widetext}
Since the lower limit in the first term of this expression vanishes in view of 
Eq. (\ref{eq:ypsikeq}),  
the integration of the second term of Eq. (\ref{eq:peq}) by parts results in:
\begin{widetext}
\begin{eqnarray}
y_{p}(z)= f_{p-1}(z)+\left(-\frac{1}{2 y_{p-1}(z)}\frac{d\,f_{p-1}(z)}{dz}\right)
-\int_{z_0}^{z}ds \frac{d}{ds}\left(-\frac{\frac{d\,f_{p-1}(s)}{ds}}{2 y_{p-1}(s)}
 \right)\;\exp[-2 \int_{s}^{z} y_{p-1}(t)) dt],  \nonumber \\
\label{eq:pi0qeq}
\end{eqnarray}
\end{widetext}
The successive integrations by parts of Eq. (\ref{eq:pi0qeq}) 
lead \cite{RV} to the series
\beq
y_{p}(z)= \sum_{n=0}^{\infty}\mathcal{L}_n^{(p)}(z)
\label{eq:piqeq}
\eeq
with $\mathcal{L}_n^{(p)}(z)$ given by recursive relation
\beq
\mathcal{L}_n^{(p)}(z)=\frac{1}{2\,y_{p-1}(z)}\frac{d}{dz}
(-\mathcal{L}_{n-1}^{(p)}(z))
\label{eq:rrqeq}
\eeq
and
\beq
\mathcal{L}_n^{(0)}(z)=f_{p-1}(z).
\label{eq:rr0qeq}
\eeq
Since 
\begin{eqnarray}
\frac{d}{dz}=g\frac{d}{dr},\;\;g=\lambda^{-1}=\frac{\hbar}{\sqrt{2 m}}, 
\label{eq:geq}
\end{eqnarray}
Eq. (\ref{eq:piqeq}) represents the expansion of the p-th QLM iterate in 
powers of $g$,  
that is in powers of $\hbar$, which one can compare with the WKB series
as will be done in the next section.

For the zeroth iterate $y_{0}(z)$ it seems natural to choose the zero WKB
approximation that is to set 
\begin{eqnarray}
y_{0}(z)=i k(z), 
\label{eq:00eq}
\end{eqnarray}
which in addition satisfies boundary condition (\ref{eq:bceq}).
However, one has to be aware that this choice has unphysical turning point
singularities. According to the existence theorem for linear
differential equations \cite{In}, if $y_{p-1}(z)$ in Eq.\ (\ref{eq:qeq}) 
is a discontinuous function of $z$ in a certain interval, then 
$y_{p}(z)$ or its derivatives may also be discontinuous functions in
this interval, so consequently the turning point singularities 
of $y_{0}(z)$ may propagate to the next iterates.\\

Substitution of the initial QLM approximation (\ref{eq:00eq})
into Eq.\ (\ref{eq:peq}) gives especially simple expression \cite{RV} 
for the first QLM iterate
\begin{eqnarray}
 y_{1}(z)&=& i k(z)-i \int_{z_0}^{z} ds\;k'(s)\;
\exp[-2 i \int_{s}^{z} k(t) dt],  \nonumber \\
\label{eq:1eq}
\end{eqnarray}

Thus the first QLM iterate is expressible in a 
closed integral form. However, it takes 
 into account, though approximately, infinite number of the WKB
 terms corresponding to higher powers of $\hbar$ as it will be shown 
 in the next section. In view of this it is a better 
 approximation than the usual WKB. 
\section{Comparison of expansions of QLM iterates and WKB series}
To obtain the WKB series one has to expand solution $y$ of the Riccati
equation (\ref{eq:weq}) in powers of $\hbar$.This is easy to do 
by using Eq. (\ref{eq:geq}) and looking for $y$ in the form of series 
in $g$:
\beq
y=\sum_{m=0}^{\infty} g^m Y_m\;.
\label{eq:sumwkb}
\eeq
Substitution into (\ref{eq:weq}) and equation of terms by 
the identical powers of $g$ gives 
\beq
\frac{d\,Y_{m-1}}{dr}=-\sum_{k=0}^{m} Y_k\;Y_{m-k}.
\label{eq:recwkb}
\eeq
This reduces to the recurrence relation
\beq
Y_{m}=-\frac{1}{2\,Y_0}\,\left(Y'_{m-1}+\sum_{k=1}^{m-1} 
Y_k\;Y_{m-k}\right).
\label{eq:redwkb}
\eeq
The derivatives in this and subsequent expressions are in variable $r$.
The zero WKB approximation $Y_0$ is given by $Y_0=i\,k$. 
The subsequent terms $Y_{n}$
of the expansion could be obtained from this recurrence relation 
by use of Mathematica \cite{Math}.

 We present here the WKB expansion (\ref{eq:sumwkb}) up to  $g^7$ 
 inclusively:
 
\begin{widetext}
\begin{eqnarray} 
  y = i \,k - \frac{g\,k'}{2\,k} + 
  \frac{i }{8\,k^3}\,g^2\,( 3\,{k'}^2 
  - 2\,k\,k'' )  
  + \frac{g^3\,}{8\,{k}^5}
     ( 6\,{k'}^3 - 6\,k\,k'\,k'' + 
       {k}^2\,k^{(3)} )+ 
  \frac{i }{128\,k^7}\,g^4\,
     ( -297\,{k'}^4 + 
       396\,k\,{k'}^2\,k'' \nonumber\\- 
       52\,{k}^2\,{k''}^2 - 
       80\,{k}^2\,k'\,k^{(3)} +
       8\,{k}^3\,k^{(4)} ) - 
  \frac{g^5\,}{32\,
     {k}^9}( 306\,{k'}^5 - 
       510\,k\,{k'}^3\,k'' + 
       111\,{k}^2\,{k'}^2\,k^{(3)} - 
       3\,{k}^2\,k'\,
        ( -48\,{k''}^2 + 
          5\,k\,k^{(4)} )\nonumber\\ + 
       {k}^3\,( -24\,k''\,k^{(3)} + 
          k\,k^{(5)} )  )   
      + \frac{i }{1024\,k^{11}}\,g^6\,
      ( 50139\,{k'}^6 - 
        100278\,k\,{k'}^4\,k''  + 
        22704\,{k}^2\,{k'}^3\,k^{(3)} + 
        12\,{k}^2\,{k'}^2\, 
         ( 3679\,{k''}^2 \nonumber\\- 
           290\,k\,k^{(4)} )  + 
        16\,{k}^3\,k'\,
         ( -694\,k''\,k^{(3)} + 
           21\,k\,k^{(5)} )  - 
        8\,{k}^3\,       
        ( 301\,{k''}^3 - 
           80\,k\,k''\,k^{(4)} + 
           k\,( -49\,{k^{(3)}}^2 + 
              2\,k\,k^{(6)} )  ) ) \nonumber\\ + 
  \frac{g^7}{128\,k^{13}} ( 38286\,{k'}^7 - 
       89334\,k\,{k'}^5\,k'' + 
       20721\,{k}^2\,{k'}^4\,k^{(3)} + 
       {k'}^3\,( 53724\,{k}^2\,{k''}^2 - 
           3405\,{k}^3\,k^{(4)} )  + 
       3\,{k}^3\,{k'}^2\,
        ( -5426\,k''\,k^{(3)} \nonumber\\+ 
          129\,k\,k^{(5)} )  + 
       2\,{k}^3\,k'\,
        ( -3528\,{k''}^3 + 
          735\,k\,k''\,k^{(4)} + 
          2\,k\,( 225\,{k^{(3)}}^2 - 
             7\,k\,k^{(6)} )  )  + 
       {k}^4\,( 1176\,{k''}^2\,k^{(3)} - 
          62\,k\,k''\,k^{(5)} \nonumber\\+ 
          k\,( -90\,k^{(3)}\,k^{(4)} + 
             k\,k^{(7)})))\nonumber\\
\label{eq:wkbeq}
\end{eqnarray}
\end{widetext}
To compare expansion of the first QLM iterate $y_1$ 
in powers of $\hbar$ 
with the WKB expansion (\ref{eq:wkbeq}) we have to use, as we 
have already mentioned in the previous section, 
Eqs. (\ref{eq:piqeq}) and (\ref{eq:rrqeq}) together with
 Eq. (\ref{eq:00eq}). The result 
up to  power $g^7$ inclusively is again obtained with 
the help of Mathematica \cite{Math}:
\begin{widetext}
\begin{eqnarray} 
y_1 = i \,k - \frac{k'\,g}{2\,k} + 
  \frac{i\, g^2}{4\,k^3}\,
     ( {k'}^2 - k\,k'' ) 
      + \frac{g^3}{8\,{k}^5}( 3\,{k'}^3 - 
       4\,k\,k'\,k'' + {k}^2\,k^{(3)}
       )  + 
  \frac{i\,g^4}{16\,{k}^7}
     ( -15\,{k'}^4 + 
       25\,k\,{k'}^2\,k'' - 
       7\,{k}^2\,k'\,k^{(3)} \nonumber\\+ 
       {k}^2\,( -4\,{k''}^2 + 
          k\,k^{(4)} )  ) 
           + \frac{g^5}{32\,{k}^9}( -105\,{k'}^5 + 
       210\,k\,{k'}^3\,k'' - 
       60\,{k}^2\,{k'}^2\,k^{(3)} + 
       {k}^2\,k'\,
        ( -70\,{k''}^2 + 
          11\,k\,k^{(4)} )\nonumber\\  + 
       {k}^3\,( 15\,k''\,k^{(3)} - 
          k\,k^{(5)} )  ) 
    - \frac{i \,g^6}{64\,{k}^{11}}\,
     ( -945\,{k'}^6 + 
       2205\,k\,{k'}^4\,k'' - 
      630\,{k}^2\,{k'}^3\,k^{(3)} + 
       14\,{k}^2\,{k'}^2\,
       ( -80\,{k''}^2 + 
         9\,k\,k^{(4)} )\nonumber\\  + 
      2\,{k}^3\,k'\,
        ( 175\,k''\,k^{(3)} - 
          8\,k\,k^{(5)} )  + 
       {k}^3\,( 70\,{k''}^3 - 
          26\,k\,k''\,k^{(4)} + 
         k\,( -15\,{k^{(3)}}^2 + 
            k\,k^{(6)} )  )  ) 
     + \frac{\,g^7}{128\,{k}^{13}}( 10395\,{k'}^7\nonumber\\ - 
      27720\,k\,{k'}^5\,k'' + 
       7875\,{k}^2\,{k'}^4\,k^{(3)} - 
       126\,{k}^2\,{k'}^3\,
        ( -150\,{k''}^2 + 
          13\,k\,k^{(4)} )  + 
      14\,{k}^3\,{k'}^2\,
       ( -495\,k''\,k^{(3)} + 
          17\,k\,k^{(5)} ) \nonumber\\ + 
       {k}^3\,k'\,
        ( -2800\,{k''}^3 + 
          784\,k\,k''\,k^{(4)} + 
         k\,( 455\,{k^{(3)}}^2 - 
            22\,k\,k^{(6)} )  )  + 
       {k}^4\,( 560\,{k''}^2\,k^{(3)} - 
         42\,k\,k''\,k^{(5)} \nonumber\\+ 
         k\,( -56\,k^{(3)}\,k^{(4)} + 
             k\,k^{(7)} )  )  ) \nonumber\\    
\label{eq:y1eq}
\end{eqnarray}
\end{widetext}
The comparison of expansion of the first QLM iterate 
in powers of $\hbar$ and  WKB series was originally 
performed in the works \cite{RV,J}. There it was shown that 
the expansion
reproduces exactly the first two terms 
and also gives correctly the structure of the WKB series up to 
the power $g^3$ considered in these works, generating  series with 
identical terms, but with different coefficients.  
Comparison of 
Eqs. (\ref {eq:wkbeq})
and (\ref{eq:y1eq} ) of the present work shows that this 
conclusion is true also
if higher powers of $g$ are taken into account. 

 The computation of the expansion of the second QLM iterate $y_{2}$
 in powers of $\hbar$
is done by reexpanding the term $\frac{1}{2\,y_{1}}$ 
in Eq. (\ref{eq:rrqeq}) in the series in powers of $g$ 
and keeping the powers up to $g^7$ inclusively in 
this expression as well as in the sum in Eq. (\ref{eq:piqeq}).
The result is given by:
\begin{widetext}
\begin{eqnarray} 
y_2 = i \,k - \frac{k'\,g}{2\,k} + 
  \frac{i \,g^2
     }{8\,{k}^3}\,
     ( 3\,{k'}^2 - 2\,k\,k'' )  + \frac{\,g^3}{8\,{k}^5}\,( 6\,{k'}^3 - 
       6\,k\,k'\,k'' + {k}^2\,k^{(3)}
       )  + 
  \frac{i\,g^4 }{32\,{k}^7}\,
     ( -74\,{k'}^4 + 
       99\,k\,{k'}^2\,k'' - 
       20\,{k}^2\,k'\,k^{(3)} \nonumber\\+ 
       {k}^2\,( -13\,{k''}^2 + 
         2\,k\,k^{(4)} )  )  
   + \frac{\,g^5}{64\,
     {k}^9}( -607\,{k'}^5 + 
       1017\,k\,{k'}^3\,k'' - 
      222\,{k}^2\,{k'}^2\,k^{(3)} + 
      6\,{k}^2\,k'\,
       ( -48\,{k''}^2 + 
          5\,k\,k^{(4)} ) \nonumber\\ - 
       2\,{k}^3\,( -24\,k''\,k^{(3)} + 
          k\,k^{(5)} )  ) 
          - \frac{i \,g^6}{128\,{k}^{11}}\,
     ( -6186\,{k'}^6 + 
      12446\,k\,{k'}^4\,k'' - 
      2832\,{k}^2\,{k'}^3\,k^{(3)} + 
       {k}^2\,{k'}^2\,
       ( -5503\,{k''}^2 + 
          435\,k\,k^{(4)} ) \nonumber\\ + 
       2\,{k}^3\,k'\,
        ( 694\,k''\,k^{(3)} - 
          21\,k\,k^{(5)} )  + 
      {k}^3\,( 301\,{k''}^3 - 
        80\,k\,k''\,k^{(4)} + 
          k\,( -49\,{k^{(3)}}^2 + 
             2\,k\,k^{(6)} )  )  ) + 
 \frac{\,g^7}{256\,{k}^{13}}( 75256\,{k'}^7\nonumber\\ - 
       176659\,k\,{k'}^5\,k'' + 
       41224\,{k}^2\,{k'}^4\,k^{(3)} + 
       4\,{k}^2\,{k'}^3\,
        ( 26687\,{k''}^2 - 
          1700\,k\,k^{(4)} )  + 
       2\,{k}^3\,{k'}^2\,
        ( -16243\,k''\,k^{(3)} + 
          387\,k\,k^{(5)} ) \nonumber\\ + 
       {k}^3\,k'\,
        ( -14071\,{k''}^3 + 
          2940\,k\,k''\,k^{(4)} + 
          8\,k\,( 225\,{k^{(3)}}^2 - 
             7\,k\,k^{(6)} )  )  + 
       2\,{k}^4\,( 1176\,{k''}^2\,
           k^{(3)} - 62\,k\,k''\,k^{(5)}\nonumber\\ + 
          k\,( -90\,k^{(3)}\,k^{(4)} + 
             k\,k^{(7)} )  )  ) \nonumber\\
\label{eq:y2eq}
\end{eqnarray}
\end{widetext}
The expansion of $y_2$ reproduces exactly already the first four terms 
of the WKB series. It also gives  the proper structure of 
the other terms of the WKB series, generating  series with 
identical terms which have approximately correct coefficients. 

The expansion of $y_3$ is obtained in the similar fashion.
 It reproduces exactly the first eight terms of the WKB series
 that is all the terms up to the power $g^7$ inclusively listed in 
 Eq. (\ref{eq:wkbeq}). 
 
 Summing up, we have proved that the expansion of the 
 first, second and third QLM iterates reproduces exactly 
 two, four and eight WKB terms respectively. Since the 
 zero QLM iterate $y_0$ was chosen to be equal to the zero WKB
 approximation $i\,k$, one can state that the p-th QLM iterate
 contains $2^p$ exact terms. In addition expansion of each QLM 
 iterate has the correct structure  whose 
terms are identical to of the WKB series with 
approximate coefficients.

The  $2^p$ law is, of course, not accidental. The QLM iterates
are quadratically convergent \cite{VBM1,VBM2,K}, that is the 
norm of the difference of the
exact solution and the p-th QLM iterate $\|y-y_p\|$ is proportional 
to the square
of the norm of the differences of the  exact solution and 
the (p-1)-th QLM iterate:
\beq
\|y-y_p\| \sim {\|y-y_{p-1}\|}^2.
\label{eq:norm}
\eeq

Here norm $\parallel g \parallel$ of function $g(x)$ 
is a maximum of the function $g(x)$ on the whole interval 
of values of $x$.

Since $y_0$ contains just one correct WKB term of power $g^0$
 and thus
$\|y-y_0\|$ is proportional to $g$, one has  to expect 
that $\|y-y_1\| \sim g^2$
and thus $y_1$ contains two correct WKB terms of powers 
$g^0$ and $g^1$. The difference $\|y-y_2\| \sim \|y-y_1\|\sim g^4$ 
so that $y_2$ contains four correct WKB terms of 
powers $g^0$, $g^1$, $g^2$  and $g^3$. Finally, the difference
$\|y-y_3\|$ should be proportional to $g^8$, and therefore
$y_3$ has to contain eight correct terms with powers 
between $g^0$ and $g^7$ inclusively. This explains the $2^p$ law.
\section{QLM and WKB energy calculations} 
The exact quantum mechanical quantization condition for 
the energy \cite{D,LP1, LP2} has the form: 

\begin{equation}
J = {\oint}_C\, y(z)dz =i\,2 \pi n .
\label{eq1}
\end{equation} 
 Here $y(z)$ is the logarithmic derivative of the wave function, given by  Eq. 
(\ref{eq:weq}), $z = gr$,  $n = 0,1,2,...$ counts the number of poles
 of $y(z)$ and is the
bound state number and the integration is along a path C in the complex
plane encircling the segment of the $\Re z$ axis between the turning points.

 The p-th QLM iterate $y_p(z)$, as we have seen, contains in addition
to $2^p$ exact WKB terms of powers $g^0, g^1,..., g^{2^p-1}$
 also an infinite number of structurally 
correct WKB terms of higher powers of $g$ with approximate coefficients.
 One can expect therefore that the quantization condition (\ref{eq1})
 with $y(z)$ approximated by any QLM iterate $y_p(z)$ including the first
\begin{equation}
J_p = {\oint}_C\, y_p(z)dz =i\,2 \pi n ,\, p=1,2,...
\label{eqp1}
\end{equation}
gives more accurate energy than the WKB quantization condition
 \beq
{\oint}_C\, k(z)dz = 2 \pi (n + \frac{1}{2}) .
\label{eq:wkbqc}
\eeq
which  is obtained by substituting into
 exact quantization condition (\ref{eq1})
the WKB expansion (\ref{eq:wkbeq}) up to the first power
 of $g \sim \hbar$,
that is $y(z)= ik(z) - \frac{\frac{d k(z)}{dz}}{2 k(z)}$,
and neglecting all higher powers of $g$ in the expansion.
Indeed, we prove in this section that Eq. (\ref{eqp1}) leads to exact 
energies not only for the the Coulomb and harmonic oscillator potentials
as it was shown earlier in ref. \cite{RV,J}, but for many other well 
known  physical potentials used in molecular and nuclear physics such as 
the P\"{o}schl-Teller, Hulthen, Hyleraas, Morse, Eckart etc.  
The WKB quantization condition (\ref{eq:wkbqc}) yields the exact energy
only for the first two potentials, but not for the rest of them.

Let us now consider the above mentioned examples:
\subsection{Harmonic oscillator 
$V(x)\,=\,\frac{x^2}{2},\,\, -\infty < x < \infty$.} 
From now on we will work in the units $\hbar=m=1$ so that from $z=\lambda x$
follows $x= \frac{z}{\sqrt{2}}$ and $V(z)=\frac{z^2}{4}$.
 
In view of the boundary condition (\ref{eq:ppeq}) $y_p(z)$ at the infinity
 should behave like $i \sqrt{E-\frac{z^2}{4}}\simeq -\frac{z}{2}+\frac{E}{z}$
where we omitted terms of  order $\frac{1}{z^2}$ and higher. Here
 we took into account that for bound states the logarithmic derivative at 
the infinity should be negative. More accurately  the pole structure of
 $y_p(z)$ at $z \sim \infty$
could be found by looking for the solution there in the form 
$y_p(z) \simeq -\frac{z}{2} +\frac{\alpha_p}{z}$. Substituting into
 the  quasilinearized equation (\ref{eq:qeq}) 
\beq 
\frac{dy_{p}(z)}{dz}=y_{p-1}^2(z)-2 y_{p}(z)y_{p-1}(z)-(E-\frac{z^2}{4})) 
\label{eq:hoqeq} 
\eeq
and again neglecting terms of order $\frac{1}{z^2}$ and higher
 which does not contribute to the integral yields $\alpha_p=E-\frac{1}{2}$, 
so that the pole term in  $y_p(z), \,\,p=1,2,...$ is given by 
$\frac{E-\frac{1}{2}}{z}$ .

The integration in Eq. (\ref{eqp1}) is counterclockwise 
along a path C in the complex
plane encircling the segments of the $\Re z$ axis between the two turning
 points $-2 \sqrt{E}$ and $2 \sqrt{E}$ . Since the only singularity
 outside contour C in the complex plane lays at infinity, the 
integral (\ref{eqp1}) could be done by distorting the contour
 to enclose the pole at $x = \infty$ . The evaluation of the integral  
 yields $i 2\pi\,(E-\frac{1}{2})= i 2\pi\, n$
or $E=n+\frac{1}{2}$, which is the exact equation for the energy levels.

\subsection{Spherical harmonic oscillator \\
$V(r)\,=\,\frac{r^2}{2} ,\,\, 0 < r < \infty$.}

This case was already considered in the work \cite{J}.
 We present here somewhat different derivation.
Eq. (\ref{eq:qeq}) now has the form 
\beq 
\frac{dy_{p}(z)}{dz}=y_{p-1}^2(z)-2 y_{p}(z)y_{p-1}(z)-(E-\frac{z^2}{4}
-\frac{l(l+1)}{z^2})) 
\label{eq:3hoqeq} 
\eeq
where $z=\sqrt{2} r$. The pole structure at  $z \sim \infty$
 as before will be given by Eq. (\ref{eq:3hoqeq}). In order to find the pole 
at $z \sim 0$ we look for $y_p$ near zero in the form $\frac{a_p}{z}+....$
where terms proportional to $\frac{1}{z^2}$, $\;\frac{1}{z^3}$ etc. are 
not explicitly displayed since they do not contribute to the integral. 
The substitution into Eq. (\ref{eq:3hoqeq}) yields 
\beq 
a_p (2 a_{p-1} - a_p -1)= l (l+1)
\label{eq:aqeq} 
\eeq
When $p$ becames large we expect that QLM iterates converge to each other 
and to the exact solution, so for large p one expects $a_{p-1}=a_p$.
The solution of Eq. (\ref{eq:aqeq}) satisfying this condition will 
be $a_p=(l+1)$, so the pole term at $z \sim 0$ is $\frac{l+1}{z}$. 
This, of course, agree also with a $z^{l+1}$ behavior of the radial 
wave function at zero, which leads to the $\frac{l+1}{z}$ behavior
 of its logarithmic derivative.
The contour C encloses two positive turning points $z_1$ and $z_2$
 defined by the equation 
\beq
E-\frac{z^2}{4}-\frac{l(l+1)}{z^2}=0
\label{eq:tpeq} 
\eeq
and the section $\Re z$ axis between them. Indeed, the physical motion takes
 place only for positive real $z$. The integrand, as we saw, 
has the poles at $z=0$ and $z=\infty$. However, in addition \cite{LP1,LP2}, 
 when
 C is distorted to enclose these poles
 there will be also
 contribution to the integral $J_p$ from the cut on the negative
  $\Re z$ axis between
 the unphysical turning points $-z_2$ and $-z_1$. This contribution
 due to the symmetry of the potential with respect to $z=0$ will be
 $-J_p$ since the path of integration around  the cut on negative
 real axis will be clockwise.
 The contribution of the poles is $2 \pi i$ multiplied by $-(l+1)$ and 
$E-\frac{1}{2}$ respectively,
so one obtains $J_p = -J_p + 2 \pi i [E-\frac{1}{2}-(l+1)]$. Thus 
$J_p= \pi i ( E-\frac{1}{2}-(l+1) ) = 2 i \pi n$ or

\beq 
 E = 2 n +l + \frac{3}{2}
\label{eq:3doeq} 
\eeq
 which is the exact equation for the energy levels of the
 three-dimensional oscillator.

\subsection{Coulomb potential $V(r) = -\frac{Z}{r} ,\,\, 0 < r < \infty$.}

This case was considered before in the work \cite{RV}.
In our derivation we set, as in the previous section  $z=\sqrt{2} r$, so 
$k^2(z)=E+\frac{Z \sqrt{2}}{z} - \frac{l (l+1)}{z^2}$ . At large z, in view
 of the boundary condition (\ref{eq:ppeq}) $y_p(z)$ at the infinity 
 should behave like $i \sqrt{-\vert E \vert+\frac{Z \sqrt{2}}{z}} \simeq 
- \sqrt{\vert E \vert} + \frac{Z}{z \sqrt{2 \vert E \vert}}$.
where we took into account that for bound states the logarithmic derivative at 
the infinity should be negative and omitted terms of order $\frac{1}{z^2}$
 and higher. The residue of the pole at the infinity thus is 
$\frac{Z}{\sqrt{2 \vert E \vert}}$. The  residue of the pole at 
$z \simeq 0$ is computed as in the previous section and yields the 
same value $l+1$.

The contour C encloses two turning
 points 
 defined by the equation 
$-\vert E \vert-\frac{Z \sqrt{2}}{z}-\frac{l(l+1)}{z^2}=0$ and
 the section $\Re z$ axis between them. This equation, unlike equation
 (\ref{eq:tpeq}) for the turning points in the previous section which
 had two positive and two negative real roots, has only two 
positive real roots $z_1 = \frac{Z}{\sqrt{2 \vert E \vert}} 
- \sqrt{\frac{Z^2}{2 E^2} - \frac{l (l+1)}{\vert E \vert}}$ and  
$z_2 =\frac{Z}{\sqrt{2 \vert E \vert}} 
+ \sqrt{\frac{Z^2}{2 E^2} - \frac{l (l+1)}{\vert E \vert}}$. Therefore
 unlike the previous case there is no cut along the negative part of
 the real axis and when C is distorted to enclose the poles 
at zero and the infinity only contribution of these poles should
 be taken into account. One obtains $ 2 \pi i 
(\frac{Z}{\sqrt{2 \vert E \vert}} - (l+1))= i 2 \pi n$ or 
$E = -\frac{Z^2}{2 (n+l+1)^2}$ which are exact energy levels
 in the Coulomb potential.\\  

\subsection{Cotangent potential\\ 
$V(x)\,=\,V_0 cot^2(\frac{\pi x}{a}) , \, V_0 > 0,\,\, 0 < x < a$.}
Let us introduce a new variable 
\beq 
z=sin^2(\frac{\pi x}{a}) 
\label{eq:sin} 
\eeq
so that $x=\frac{a}{\pi} arcsin \sqrt{z}$ ,  
$dx=\frac{a}{2 \pi} \frac{1}{\sqrt{z (1-z)}} dz$ and 
$k(z) = \sqrt{2 (E+V_0) - \frac{V_0}{z}}$.
The QLM equation (\ref{eq:qeq}) will now have a form 
\beq 
\frac{2 \pi}{a} \sqrt{(1-z) z} \frac {d y_p(z)}{dz} = 
y^2_{p-1}(z) -2  y_p(z) y_{p-1}(z) - k^2(z) .
\label{eq:sineq} 
\eeq
One of the singularities of $k^2(z)$ is at $z = 0$. Near this point
 the equation 
has a form  
\beq 
\frac{2 \pi}{a} \sqrt{z} \frac {d y_p(z)}{dz} = 
y^2_{p-1}(z) -2  y_p(z) y_{p-1}(z) + \frac{2 V_0}{z} .
\label{eq:sinz0eq} 
\eeq
We look for solution of this equation in a form $ y_p(z) = 
\frac{a_p}{\sqrt{z}}$.
Then we obtain for $a_p$ the following recurrence relation:
\beq 
a_p (2 a_{p-1} - \frac{\pi}{a})= a^2_{p-1} +
 (\frac{\pi}{a})^2 \lambda (\lambda - 1).
\label{eq:z0eq} 
\eeq
Here we set  $2 V_0 = (\frac{\pi}{a})^2 \lambda (\lambda - 1)$ where
$\lambda = \frac{1}{2} + \sqrt{\frac{1}{4} + \frac{2 V_0 a^2}{\pi^2}}$.
The solution of this algebraic equation which fullfils the demand that
 at large p
$a_p = a_{p-1}$ is $a_p =\frac{\pi}{a} \lambda $. The $y_p(z)$ near zero
thus has a form   $ y_p(z) = \frac{\pi \lambda}{a \sqrt{z}}$.
 At $z \simeq \infty$ Eq. \ref{eq:sineq} reduces to 
$\frac{2 \pi}{a} \sqrt{-z^2} \frac {d y_p(z)}{dz} = 
y^2_{p-1}(z) -2  y_p(z) y_{p-1}(z) - 2 (E+V_0)$. Looking for solution
in a form $y_p(z)=c_p$ where $c_p$ is some constant, one obtains
 the recurrent relation for $c_p$, namely 
$c^2_{p-1} - 2 c_p c_{p-1} - 2 (E+V_0)=0$. The solution 
of this algebraic equation which fullfils the demand that at large p
$c_p = c_{p-1}$ is $c_p = \sqrt{ - 2 (E+V_0)}$.

The quantization condition (\ref{eqp1}) in variable z given by 
Eq. (\ref{eq:sin}) has a form 
\begin{equation}
J_p =\frac{a}{\pi} {\oint}_C\,\frac{y_p(z)}{\sqrt{z (1-z)}} 
dz =i\,2 \pi n ,\, p=1,2,...
\label{eq:zpeq} 
\eeq
where integration is  counterclockwise along a path C in the complex
plane encircling the cut  along the $\Re z$ axis between the zero and 
$z= \frac{V_0}{2 (E + V_0)}$. Since $y_p(z)$ equals to 
$\frac{\pi \lambda}{a \sqrt{z}}$ and $\sqrt{- 2 (E+V_0)}$ at $z \simeq 0$
and $z \simeq \infty$ respectively, the integrand in Eq. (\ref{eq:zpeq})
has poles with residues $\frac{\pi \lambda}{a}$ and $\sqrt{2 (E+V_0)}$ there.
The deformation of the contour to include these poles and computation
 of their contributions yields
\begin{equation}
J_p =2 \pi i (-\frac{a}{\pi} \lambda +\sqrt{2 (E+V_0)}) =2 \pi i n ,\,
\label{eq:jzpeq} 
\eeq
or, upon substitution of value of $\lambda$ and inserting $\hbar$
and $m$ from dimensional considerations,\\ 
$E=-V_0 +\frac{\pi^2 \hbar^2}{2 m a^2}(n + \frac{1}{2} + 
\sqrt{\frac{2 m V_0 a^2}{\pi^2 \hbar^2} + \frac{1}{4}})$.\\
This is the exact equation for the energy levels in the cotangent potential 
\cite{GK}.
The correspondent WKB expression is different and given by  \cite{GK}\\
$E=-V_0 +\frac{\pi^2 \hbar^2}{2 m a^2}(n + \frac{1}{2} + 
\sqrt{\frac{2 m V_0 a^2}{\pi^2 \hbar^2}})^2$.

\subsection{P\"{o}schl-Teller potential hole\\ 
$V(x)\,=\,\frac{V_1}{sin^2(\frac{\pi x}{a})} + 
\frac{V_2}{cos^2(\frac{\pi x}{a})} , \, V_1 > 0, V_2 > 0,\,\, 
0 < x < \frac{a}{2}$.}
This potential is a generalization of  the cotangent potential 
 and reduces to it in case $V_1=V_2 \equiv \frac{V_0}{4}$.
Indeed, in this case 
$V=\frac{V_0}{sin^2(\frac{\pi y}{a})}=V_0 [( cot^2(\frac{\pi y}{a})+1]$
where $y\equiv 2 x$ changes between $0$ and $a$.
The computation therefore proceeds as in the previous section
with a difference that in addition to the poles at zero and the infinity 
there is also a pole at $z =1$. Indeed, using again variable $z$ defined by
 Eq. (\ref{eq:sin}) one writes the Eq. (\ref{eq:qeq}) in a form 
 (\ref{eq:sineq}) with  $k^2(z)$ now given by $2 (E - \frac{V_1}{z}
 -\frac{V_1}{1-z})$.  

The poles at $z=0$, $z=1$ and $z=\infty$ are computed as in the previous
 section and have respectively the form $\frac{\pi \lambda_1}{a z}$,
  $\frac{\pi \lambda_2}{a (z-1)}$ and $\frac{\sqrt{2 E}}{z}$. Here
$\lambda_k = \frac{1}{2} + \sqrt{\frac{1}{4} + \frac{2 V_k a^2}{\pi^2}}$,
 so that  $2 V_k = (\frac{\pi}{a})^2 \lambda_k (\lambda_k - 1)$;  $k=1,2$.
The  integration in the quantization condition (\ref{eqp1} is
  counterclockwise along a path C in the complex
plane encircling the cut  along the $\Re z$ axis between the two
 turning points $z_1, z_2$ given by $z_{1,2}=
\frac{1}{2 E}[(E+V_1 -V_2) \pm \sqrt{(E+V_1 -V_2)^2 - 4 E V_1}]$ .
The deformation of the contour to include the poles and computation
 of their contributions yields
\begin{equation}
J_p =2 \pi i (-\frac{a}{\pi} \lambda_1 - \frac{a}{\pi} \lambda_2
 +\sqrt{2 E}) =2 \pi i n ,\,
\label{eq:jjzpeq} 
\eeq
or, upon substitution of values of $\lambda_k$ and inserting $\hbar$
and $m$ from dimensional considerations,\\ 
$E=\frac{\pi^2 \hbar^2}{2 m a^2} [(2 n + 1 + 
\sqrt{\frac{2 m V_1 a^2}{\pi^2 \hbar^2} + \frac{1}{4}} + 
\sqrt{\frac{2 m V_2 a^2}{\pi^2 \hbar^2} + \frac{1}{4}})^2$.\\
This is the exact equation for the energy levels in the
 P\"{o}schl-Teller potential hole \cite{F}.
The correspondent WKB expression is different and given by\\
$E=\frac{\pi^2 \hbar^2}{2 m a^2} [2 n + 1 + 
\sqrt{\frac{2 m V_1 a^2}{\pi^2 \hbar^2}} + 
\sqrt{\frac{2 m V_2 a^2}{\pi^2 \hbar^2}}]^2$.\\

\subsection{Modified P\"{o}schl-Teller potential\\ 
$V(x)\,=\,-\frac{V_0}{cosh^2(\frac{x}{a})} , \, V_0 > 0, \,\, 
- \infty < x < \infty$.}
Setting $z=cosh^2(\frac{x}{a})$ so that $dx =\frac{a}{2 z (z-1)} dz$ 
one obtains that the  QLM equation (\ref{eq:qeq}) now has a form 
\beq 
\frac{2}{a} \sqrt{(z-1) z} \frac {d y_p(z)}{dz} = 
y^2_{p-1}(z) -2  y_p(z) y_{p-1}(z) - k^2(z) .
\label{eq:mpteq} 
\eeq
where  $k^2(z)$ is given by $2 (-\vert E \vert  + \frac{V_0}{z})$. 
At $z \simeq 0$ this equation has a form 
\beq 
\frac{2 \pi}{a} \sqrt{- z} \frac {d y_p(z)}{dz} = y^2_{p-1}(z)
 - 2  y_p(z) y_{p-1}(z) -  \frac{\lambda (\lambda-1)}{z}
\label{eq:0mpteq} 
\eeq
 where we use the definition of 
$\lambda = \frac{1}{2} + \sqrt{\frac{1}{4} + 2 V_0 a^2}$,
 so that  $2 V_0 = (\frac{\lambda (\lambda - 1)}{a^2})$.
 The solution near zero could be looked for in analogy 
with previous cases in the form $y_p(z) = \frac{a_p}{\sqrt{-z}}$.
 Substitution into (\ref{eq:0mpteq}) gives 
$a_p = a^2_{p-1} - 2 a_p a_{p-1} +\lambda (\lambda-1)$
whose solution, satisfying the condition that at large $p$ will be
 $ a_p= a_{p-1}$ is $ a_p = \lambda - 1$. Thus near zero 
$y_p(z) \simeq  \frac{\lambda-1}{\sqrt{-z}}$. The solution
at the infinity $y_p(z) \simeq \sqrt{2 \vert E \vert}$ is 
found in the same way as in the previous two sections.
The  integration in the quantization condition
\begin{equation}
J_p =a {\oint}_C\,\frac{y_p(z)}{\sqrt{z (z-1)}} 
dz =i\,2 \pi n ,\, p=1,2,...
\label{eq:zzpeq} 
\eeq
 is counterclockwise along a path C in the complex
plane encircling the cut  along the $\Re z$ axis between 
$z=\frac{V_0}{\vert E \vert}\equiv
\frac{\lambda (\lambda - 1)}{2 a^2 \vert E \vert}$ and $z = \infty$.
The deformation of the contour and computation of the integral
upon substitution of values of $\lambda$ and insertion of $\hbar$
and $m$ from dimensional considerations yields

\begin{eqnarray}  
E=-\frac{\hbar^2}{2 m a^2} [ - (n + \frac{1}{2}) + \nonumber
\sqrt{\frac{2 m V_0 a^2}{\hbar^2} + \frac{1}{4}}\,\,]^2.\\
\label{eq:enpeq}
\end{eqnarray}
This is the exact equation for the energy levels in the
 P\"{o}schl-Teller potential hole \cite{GK,F}.
The correspondent WKB expression is different and given by \cite{GK}\\
$E=-\frac{\hbar^2}{2 m a^2} [ - (n + \frac{1}{2}) + 
\sqrt{\frac{2 m V_0 a^2}{\hbar^2}}\,\,]^2$.\\

\subsection{Hylleraas potential\\ 
$V(x)\,=\,-\frac{V_0}{cosh^2(\frac{r}{a})} , \, V_0 > 0, \,\, 
0 < r < \infty$.}

This potential, used in the molecular and nuclear physics,
 was introduced in work \cite{Hy}. The computation 
proceeds as in previous section with the difference that now  
the physical motion takes place only for positive real $r$ and 
the wave function equals zero at the origin. The latter in view of 
the simmetry of the Hylleraas potential toward exchange $r$ to 
$-r$ means that the solution of the present problem is equivalent
to considering
 uneven states in the potential of the previous section. 
Thus in Eq. (\ref{eq:enpeq}) n should be 
changed to $(2 n - 1), n=1,2,.. $ .Therefore the energy levels in the
 Hylleraas potential are given by
\begin{eqnarray}  
E=-\frac{\hbar^2}{2 m a^2} [ - (2 n - \frac{1}{2}) + \nonumber
\sqrt{\frac{2 m V_0 a^2}{\hbar^2} + \frac{1}{4}}]^2.\\
\label{eq:enonpeq}
\end{eqnarray}
This coincides with the exact expression \cite{G} for 
the energy levels. The correspondent WKB 
expression is rather different \cite{G}: \\
$E=-\frac{\hbar^2}{2 m a^2} [ - (2 n -1) + 
\sqrt{\frac{2 m V_0 a^2}{\hbar^2}}]^2$, n=1,2,...  .\\

\subsection{Eckart potential  
$V(x)\,=-\,A\frac{e^{-\frac{x}{a}}}{1+e^{-\frac{x}{a}}}
 - B \frac{e^{-\frac{x}{a}}}{(1+e^{-\frac{x}{a})^2}}, \, B > |A|, \,\, 
-\infty < x < \infty$.}
This potential, introduced in work \cite{Eck}, like the Hylleraas 
potential is also widely 
used in molecular and nuclear physics. The potential 
$V(x) \rightarrow -A$ as $x \rightarrow -\infty$ and 
$V(x) \rightarrow 0$ as $x \rightarrow \infty$. The minimum
value $V_{min}$ of this potential is  $V_{min}=-\frac{(A+B)^2}{4 B}$;
the discrete spectrum of energy lies respectively in the interval
$(-\frac{(A+B)^2}{4 B}, min(0, -A) )$. 

Let us start from calculating
the energy levels $\epsilon=-E$ in the WKB approximation. 
The WKB quantization condition
 (\ref{eq:wkbqc}) now has the form
 \begin{eqnarray}
 \frac{\sqrt{2m}}{\hbar}{\oint}_C\, \sqrt{-\epsilon+ \nonumber
 \,A\frac{e^{-\frac{x}{a}}}{1+e^{-\frac{x}{a}}} \nonumber
 + B \frac{e^{-\frac{x}{a}}}{(1+e^{-\frac{x}{a})^2}}}dx \\
 = 2 \pi (n + \frac{1}{2})\, . 
\label{eq:ecpot}
\end{eqnarray}
Introducing a new variable $t=-e^{-\frac{x}{a}}, -\infty < t < 0$ 
reduces the WKB quantization condition (\ref{eq:ecpot}) to

\begin{eqnarray}
 -\frac{\sqrt{2 m a^2}}{\hbar}{\oint}_C\, \sqrt{-\epsilon- 
 \,A\frac{t}{1-t} \nonumber
 - B \frac{t}{(1-t)^2}} \frac{dt}{t}\\ 
 = 2 \pi (n + \frac{1}{2})\, . 
\label{eq:ecpot1}
\end{eqnarray}

The contour C encloses two turning points $t_{1,2}=
\frac{1}{2 (\epsilon - A)} [(A + B - 2\epsilon) \pm
\sqrt{(A + B)^2 -4 \epsilon B}]$ defined by the equation 
$\epsilon + 
 \,A\frac{t}{1-t} 
 + B \frac{t}{(1-t)^2}=0$ and
 the section $\Re t$ of the axis between them. 
 When C is distorted it encloses the poles 
at zero, at one and the infinity so the contributions of 
these poles with residues $\sqrt{- \epsilon} , \sqrt{- B}$ and 
$ -\sqrt{A-\epsilon}$, respectively, should be taken into account.
Since the poles at $t=0$ and $t=1$ are enencircled clockwise that 
is in opposite direction compare with
the pole at $t=\infty$ which is enencircled counterclockwise, their 
contribution enters with opposite signs and equals to 
$2 \pi i (\sqrt{-\epsilon} + \sqrt{A-\epsilon} - \sqrt{- B})$
which yields to the following 
equation for the WKB energy levels:

\begin{eqnarray}
- \frac{\sqrt{2 m a^2}}{\hbar}(\sqrt{\epsilon} + 
\sqrt{\epsilon -A} - \sqrt{B}) = (n+\frac{1}{2})
\label{eq:ecen}
\end{eqnarray}
This equation coincides with given in work \cite{G}.

To compute energy levels in the quasilinear approximation we have to use 
the  QLM equation (\ref{eq:qeq}) which after switching to variable $t$
has a form

\begin{eqnarray}
- \frac{t}{a} \frac{dy_p}{dt} = y^2_p -2 y_p\,y_{p-1}
 +\epsilon+ A \frac{t}{1-t} + B \frac{t}{(1-t)^2}\nonumber\\
\label{eq:ecenqlm}
\end{eqnarray}
For convenience of further computations we set here $\frac{2m}{\hbar^2}$ 
equal to unity. 
 The quantization condition (\ref{eqp1}) in variable $t$ is given by
\begin{equation}
J_p =-a {\oint}_C\,\frac{y_p(t)}{t} dt =i\,2 \pi n ,\, p=1,2,...
\label{eq:zpeq5} 
\eeq

At the singular point $t \sim 0 $ of the integrand of Eq. (\ref{eq:zpeq5})
 Eq. (\ref{eq:ecenqlm}) reduces to
\begin{eqnarray}
- \frac{t}{a} \frac{dy_p}{dt} = y^2_p -2 y_p\,y_{p-1}
 +\epsilon
\label{eq:ecenqlm1}
\end{eqnarray}
whose solution is $ y_p = c_p$, where  $c_p$ is a constant satisfying
the algebraic equations $c^2_p -2 c_p\,c_{p-1}+\epsilon$. Since at 
large $p$  we expect $ y_{p-1} \rightarrow y_p \rightarrow y$, where 
y is an exact solution at $t=0$, it 
means , in view of $c_p$ being a constant, that should be 
$c_{p} = c_{p-1}=c$, that is we are looking for a fixed point 
solution of Eq. (\ref{eq:ecenqlm1}) which is $c_p=\sqrt{\epsilon}$.
The positive sign before the root is chosen since  
the first term in expansion of $y_p(t)$ in the WKB terms is $ik(t)$. 
Thus  $y_p(t) \simeq i k(t)=\sqrt{\epsilon+ 
 \,A\frac{t}{1-t} + B \frac{t}{(1-t)^2}}$ that is  
 $y_p(0) \simeq +\sqrt{\epsilon} $. 

Another singular point of the integrand of Eq. (\ref{eq:zpeq5})
lays at $t \sim \infty$, since change of variable $v=\frac{1}{t}, 
t \rightarrow \infty$ when $v \rightarrow 0$, converts $\frac{y_p(t)}{t} dt$
into $-\frac{y_p(v)}{v} dv$ which has pole at $v=0$. At $t \sim \infty$
Eq. (\ref{eq:ecenqlm}) reduces to 
\begin{eqnarray}
- \frac{t}{a} \frac{dy_p}{dt} = y^2_p -2 y_p\,y_{p-1}
 +\epsilon- A 
\label{eq:ecenqlm2}
\end{eqnarray}
The solution of this equation one can look in the form $y_p=d_p$ where 
$d_p$ is a constant which satisfies the algebraic equation $d^2_p -
2 d_p d_{p-1} +\epsilon -A=0$. The fixed point $d_p=d$ of this equation is
$d=\sqrt{\epsilon -A}$ so that the residue of the integrand of 
Eq. (\ref{eq:zpeq5}) at  
$t \sim \infty$ equals $\sqrt{\epsilon -A}$.

Eq. (\ref{eq:ecenqlm}) near its singular point $t=1$ has a form
\begin{eqnarray}
- \frac{1}{a} \frac{dy_p}{dt} = y^2_p -2 y_p\,y_{p-1}
 + A \frac{1}{1-t} + B \frac{1}{(1-t)^2}\nonumber\\
\label{eq:ecenqlm3}
\end{eqnarray}
Looking for solution of this equation in the form $y_p=\frac{b_p}{t-1}$
we obtain for constants  $b_p$ the recurrence relations 
$\frac{1}{a} b_p= b^2_p -
2 b_p b_{p-1} +B$ whose solution at the fixed point $b_p= b$ of this 
equation is given by $b=-\frac{1}{2a} (1 - \sqrt{1+4 a^2 B})$ 
and $y_p \sim \frac{b}{t-1}$ at $t \sim 1$. Again, 
the positive sign before the root is chosen from the same consideration 
as in the previous paragraphs namely since in zero WKB approximation 
the residue of $y_p(t)$ at $t \sim 1$ should be of order $+\sqrt{B}$.

Summing up, the integrand of Eq. (\ref{eq:zpeq5}) has the 
three residues $\sqrt{\epsilon},\sqrt{\epsilon-A}$ and 
$-\frac{1}{2a} (1 - \sqrt{1+4 a^2 B})$ at $t=0$, 
$t=\infty$ and $t=1$ respectively.
After the reinstatement of the factor $\frac{2m}{\hbar^2}$ 
 and taking 
into account that poles $t=0$ and $=1$ are encircled in the 
opposite direction compare with the pole at the infinity (which is
encircled counterclockwise) the expression (\ref{eq:zpeq5}) 
gives therefore
\begin{eqnarray}
\sqrt{\frac{2m a^2}{\hbar^2}} (-\sqrt{\epsilon}-\sqrt{\epsilon-A})
-\frac{1}{2} (1 - \sqrt{1+\frac{8 m a^2 B}{\hbar^2}}) = n \nonumber\\
\label{eq:ecenqlm5}
\end{eqnarray}
This expression coincides with the exact expression for the Eckart
potential given in ref. \cite{G} and is different from the WKB 
quantization condition (\ref{eq:ecen}) for the energy 
obtained earlier in this section.

\subsection{Three dimensional S-wave Eckart potential 
$V(x)\,=\,-\lambda\frac{e^{-\frac{r}{a}}}{1-e^{-\frac{r}{a}}}
 + b \frac{e^{-\frac{r}{a}}}{(1-e^{-\frac{r}{a})^2}}, \, \lambda,b > 0, \,\, 
0 < r < \infty$.}

This potential is considered in works \cite{Bos,Ros,Rom} where exact and 
WKB expressions for the energy levels were obtained. Introduction of 
 a new variable $t=e^{-\frac{r}{a}}, 0 < t < 1$ 
reduces the WKB quantization condition (\ref{eq:wkbqc}) to

\begin{eqnarray}
 \frac{\sqrt{2 m a^2}}{\hbar}{\oint}_C\, \sqrt{-\epsilon+ 
 \,\lambda\frac{t}{1-t} \nonumber
 - b \frac{t}{(1-t)^2}} \frac{dt}{t}\\ 
 = 2 \pi (n + \frac{1}{2})\, . 
\label{eq:ecpot2}
\end{eqnarray}
The integrand has poles at $t=0, t=1$ and $t=\infty$ with the residues
$\sqrt{-\epsilon}, \sqrt{-b}$ and $\sqrt{-\epsilon - \lambda}$, respectively.
The calculation of the integral  therefore gives \cite{Rom}
$ 2 \pi i (- \sqrt{-\epsilon} - \sqrt{-b} + \sqrt{-\epsilon - \lambda}$ so that
the WKB energy levels could be computed from expression 

\begin{eqnarray}
\sqrt{\frac{2m a^2}{\hbar^2}} (-\sqrt{\epsilon}+\sqrt{\epsilon+\lambda}
-\sqrt{b}) = n +\frac{1}{2}\nonumber\\
\label{eq:ecenqlm6}
\end{eqnarray}
On the other side, computation of the energy levels in the quasilinear 
approximation, following the steps, outlined in previous sections leads 
to the expression

\begin{eqnarray}
\sqrt{\frac{2m a^2}{\hbar^2}} (-\sqrt{\epsilon}+\sqrt{\epsilon+\lambda})
-\frac{1}{2}\sqrt{1+\frac{8 m b}{\hbar^2}} = n +\frac{1}{2}\nonumber\\
\label{eq:ecenqlm7}
\end{eqnarray}
which is different from the WKB expression (\ref{eq:ecenqlm6}) and 
coincides with the exact expression for the energy levels, calculated 
in work \cite{Rom}.

\subsection{Three dimensional S-wave Hulthen potential 
$V(x)\,=\,-\lambda \frac{e^{-\frac{r}{a}}}{1-e^{-\frac{r}{a}}}, 
\, A > 0, \,\, 0 < r < \infty$.}

This potential is used in the nuclear physics and is a special 
case of the Eckart potential with $b=0$. The Eqs. (\ref{eq:ecenqlm6})
and (\ref{eq:ecenqlm7}) for the WKB and QLM energy levels respectively 
degenerate to 

\begin{eqnarray}
\sqrt{\frac{2m a^2}{\hbar^2}} (-\sqrt{\epsilon}+\sqrt{\epsilon+
\lambda}) = n +\frac{1}{2}\nonumber\\
\label{eq:hulecenqlm6}
\end{eqnarray}

and 

\begin{eqnarray}
\sqrt{\frac{2m a^2}{\hbar^2}} (-\sqrt{\epsilon}+\sqrt{\epsilon+\lambda})
-\frac{1}{2} = n +\frac{1}{2}\nonumber\\
\label{eq:hulecenqlm7}
\end{eqnarray}
so that the WKB and QLM energy eigenvalues are explicitly given by somewhat 
different expressions

\begin{eqnarray}
E_n = -\frac{\hbar^2}{2 m} \frac{((n+\frac{1}{2})^2-\frac{2 m a^2 \lambda}
{\hbar^2})^2}{4 a^2(n+\frac{1}{2})^2}
\label{eq:wkbhulecenqlm6}
\end{eqnarray}

and 

\begin{eqnarray}
E_n = -\frac{\hbar^2}{2 m} \frac{((n+1)^2-\frac{2 m a^2 \lambda}{\hbar^2})^2}
{4 a^2(n+1)^2}
\label{eq:qlmhulecenqlm7}
\end{eqnarray}

Here $n=0,1,2,..$. The last expression coincides with the exact 
expression for the energy levels calculated in works \cite{F,G}.
   
\subsection{One dimensional Morse potential  
$V(x)\,=\,A e^{-2 \frac{x}{a}} - B  e^{-\frac{x}{a}}, \, A,B > 0, \,\, 
-\infty < x < \infty$.}

This potential, introduced in work \cite{Mor} is heavily used 
in molecular physics computations to describe interactions 
between two molecules and for description of vibrations 
of two-atomic molecules. Introduction of 
 a new variable $t=e^{\frac{x}{a}}, 0 < t < \infty$ 
changes the WKB quantization condition (\ref{eq:wkbqc}) to

\begin{eqnarray}
 \frac{\sqrt{2 m a^2}}{\hbar}{\oint}_C\, \sqrt{-\epsilon- 
 \,\frac{A}{t^2} \nonumber
 +  \frac{B}{t}} \,\,\frac{dt}{t}\\ 
 = 2 \pi (n + \frac{1}{2})\, . 
\label{eq:ecpot3}
\end{eqnarray}
The integrand has poles at 
$t=\infty$ and at $t=0$ with the residues
$\sqrt{-\epsilon}$ and $\frac{B}{2 \sqrt{-A}}$, respectively.
The calculation of the integral  therefore gives \cite{Rom}
\begin{eqnarray}
\frac{\sqrt{2 m a^2}}{\hbar} (\sqrt{\epsilon} + \frac{B}{2 \sqrt{A}})
 = (n + \frac{1}{2})
 \label{eq:qlmhulecenqlm18}
\end{eqnarray}
which yields the well known expression \cite{F,Ros,Rom} 

\begin{eqnarray}
E_n = -\frac{B}{2} (\sqrt{\frac{B}{2 A}} - \frac{\hbar 
 (n+\frac{1}{2})}{a \sqrt{B m}})^2
\label{eq:qlmhulecenqlm8}
\end{eqnarray}
for the energy levels.

To compute energy levels in the quasilinear approximation 
one has to use the  QLM equation (\ref{eq:qeq}) which 
in variable $t$ has a form

\begin{eqnarray}
\frac{t}{a} \frac{dy_p}{dt} = y^2_p -2 y_p\,y_{p-1}+\epsilon+ 
 \,\frac{A}{t^2} - \frac{B}{t} \nonumber\\
\label{eq:ecenqlm10}
\end{eqnarray}
We set here and further $\frac{2m}{\hbar^2}$ equal to unity. 
 The quantization condition (\ref{eqp1}) in variable $t$ is given by
\begin{equation}
J_p =a {\oint}_C\,\frac{y_p(t)}{t} dt =i\,2 \pi n ,\, p=1,2,...
\label{eq:zpeq15} 
\eeq

At the singular point $t \sim \infty $ of the integrand of Eq. (\ref{eq:zpeq5})
 Eq. (\ref{eq:ecenqlm}) reduces to
\begin{eqnarray}
\frac{t}{a} \frac{dy_p}{dt} = y^2_p -2 y_p\,y_{p-1}
 +\epsilon
\label{eq:ecenqlm19}
\end{eqnarray}
whose solution is $ y_p = c_p$, where  $c_p$ is a constant satisfying
the algebraic equations $c^2_p -2 c_p\,c_{p-1}+\epsilon$. Since at 
large $p$  we expect $ y_{p-1} \rightarrow y_p \rightarrow y$, where 
y is an exact solution at $t=0$, it 
means , in view of $c_p$ being a constant, that should be 
$c_{p} = c_{p-1}=c$, that is we are looking for a fixed point 
solution of Eq. (\ref{eq:ecenqlm1}) which is $c_p=\sqrt{\epsilon}$.
As in previous paragraphs, the positive sign before the root 
is chosen since  
the first term in expansion of $y_p(t)$ in the WKB terms is $ik(t)$. 
Thus  $y_p(t) \simeq i k(t)=\sqrt{\epsilon+ 
 \frac{A}{t^2} - \frac{B}{t}}$ that is  
 $y_p(\infty) \simeq +\sqrt{\epsilon} $. 

Another singular point of the integrand of Eq. (\ref{eq:zpeq5})
lays at $t \sim 0$, where Eq. (\ref{eq:ecenqlm10}) reduces to

\begin{eqnarray}
\frac{t}{a} \frac{dy_p}{dt} = y^2_p -2 y_p\,y_{p-1}+ 
 \,\frac{A}{t^2} - \frac{B}{t} \nonumber\\
\label{eq:ecenqlm11}
\end{eqnarray}

The solution of this equation near $t=0$ has a 
form $y_p=\frac{b_p}{t} + d_p$ 
where $a_p$ and $b_p$ are constants satisfying the algebraic 
equations 
\begin{eqnarray}
b^2_p -2 b_p b_{p-1} +A = 0,\\
-\frac{b_p}{a}= 2 b_p d_p -2 (b_p d_{p-1} +b_{p-1} d_p) - B .
\label{eq:ecenqlm12}
\end{eqnarray}
 
The fixed points $b_p =b_{p-1}$ and $d_p=d_{p-1}=d$ of this equation 
are $b=\sqrt{A}$ and $d= \frac{1}{2 a} - \frac{B}{2 \sqrt{A}}$.
 From all this  the residues of the integrand at $t=\infty$ and $t=0$
 are $\sqrt{\epsilon}$ and $ \frac{1}{2 a} - \frac{B}{2 \sqrt{A}}$
 respectively. Taking 
into account that the pole at $t=0$ is encircled in the 
opposite direction compare with the pole at the infinity (which is
encircled counterclockwise)
 the expression (\ref{eq:ecenqlm10}) 
gives therefore
\begin{eqnarray}
a (\sqrt{\epsilon}-(\frac{1}{2 a} - \frac{B}{2 \sqrt{A}}) = n \nonumber\\
\label{eq:ecenqlm15}
\end{eqnarray}
This expression after reinstating the factor $\frac{2m}{\hbar^2}$
 coincides with (\ref{eq:qlmhulecenqlm18}) which means that 
 the QLM expression for the energy levels in the Morse
potential coincides with the expression (\ref{eq:qlmhulecenqlm8}) 
for the WKB energy levels obtained earlier in this section. It 
coincides also with the exact expression for the energy levels, 
see, for example, ref. \cite{LL}.
  The coincidence of the exact and the WKB eigenvalues 
  for the Morse potential is consequence of the fact that the
introduction of 
 a variable $u=\frac{2 a \sqrt{2 m A}}{\hbar}
  e^{-\frac{x}{a}}, 0 < u < \infty$ 
reduces the Schr\"{o}dinger equation with this potential
to the radial Coulomb Schr\"{o}dinger equation \cite{LL}
which, as it is well known, yields exact energy levels 
also in the WKB approximation.

\section{Conclusion}
We have shown that the quasilinearization method (QLM) which 
approaches solution of the Riccati-Schr\"{o}dinger equation 
 by approximating the nonlinear terms 
by a sequence of the linear ones, and is not based
on the existence of a smallness parameter, sums the WKB 
series. 

The advantage of the quasilinear approach
 is  that each p-th QLM iteration is 
expressible in a closed integral form. We have proved that its expansion 
in powers of $\hbar$ reproduces the structure of the WKB 
series generating an infinite number of the WKB term 
with $2^p$ terms of the expansion reproduced 
exactly and a similar number approximately. As a result one expects
 that the exact quantization condition (\ref{eq1}) with integrand
replaced by any QLM iterate (\ref{eqp1}) including the first
gives more accurate energy than the WKB quantization condition (\ref{eq:wkbqc})
  which  is obtained by substituting into exact quantization condition of 
 the WKB expansion up to the first power of $\hbar$
and neglecting all higher powers of $\hbar$. 
We show on many examples that it is indeed so and the approximation
 by QLM iterates leads to exact energies for many well 
known  physical potentials with such as the Coulomb, harmonic 
oscillator, P\"{o}schl-Teller, Hulthen, Hyleraas, Morse, Eckart etc. 

\section{Acknowledgments} 
 
The research was supported by the Israeli Science Foundation grant 131/00.

\end{document}